\documentstyle[12pt,epsfig]{article}

\newcommand{\gsim}{\raisebox{-0.13cm}{~\shortstack{$>$ \\[-0.07cm] $\sim$}}~}

\newcommand{\ra}{\rightarrow}
\newcommand{\ee}{e^+e^-}
\newcommand{\tb}{\tan \beta}
\newcommand{\s}{\smallskip}

\newcommand{\non}{\nonumber}
\newcommand{\beq}{\begin{eqnarray}}
\newcommand{\eeq}{\end{eqnarray}}

\catcode`@=11
\def\citer{\@ifnextchar
[{\@tempswatrue\@citexr}{\@tempswafalse\@citexr[]}}
\def\@citexr[#1]#2{\if@filesw\immediate\write\@auxout{\string\citation{#2}}\fi
  \def\@citea{}\@cite{\@for\@citeb:=#2\do
    {\@citea\def\@citea{--\penalty\@m}\@ifundefined
       {b@\@citeb}{{\bf ?}\@warning
       {Citation `\@citeb' on page \thepage \space undefined}}%
\hbox{\csname b@\@citeb\endcsname}}}{#1}}
\catcode`@=12

\begin{document}
\baselineskip=16.5pt

\hspace{12.1cm}   DESY 04-241  

\hspace{11.8cm}   December 2004  \\

\begin{center}

\vspace*{.7cm}

{\large\sc {\bf Prospects of mass measurements for neutral MSSM Higgs }}

\vspace*{0.3cm}

{\large\sc {\bf bosons in the intense--coupling regime at a Linear Collider}}

\vspace{0.9cm}

{\sc E. BOOS}$^{1,2}$, {\sc V. BUNICHEV}$^{1}$,  {\sc  A. DJOUADI}$^{3,4}$ 
and {\sc  H.J. SCHREIBER}$^{5}$

\vspace*{0.7cm}

\begin{small}
$^1$  Skobeltsyn Institute of Nuclear Physics, MSU, 119992 Moscow, Russia.
\vspace*{2mm}

$^2$ Fermilab, P.O. Box 500, Batavia, IL 60510-0500, USA
\vspace*{2mm}

$^3$ LPT, Universit\'e Paris--Sud \& UMR8627--CNRS, Bt. 210, F--91405 Orsay, France. 
\vspace*{2mm}

$^3$ LPMT, Universit\'e de Montpellier II, F--34095 Montpellier Cedex 5, 
France. 
\vspace*{2mm}

$^{5}$ DESY, Deutsches Elektronen-Synchrotron, D-15738  Zeuthen, Germany.
\vspace*{2mm}

\end{small}
\end{center} 

\vspace*{7mm} 

\begin{abstract} 

We analyze the prospects for detecting the three neutral Higgs bosons of the
Minimal Supersymmetric extension of the Standard Model in the intense--coupling
regime at $\ee$ colliders. Due to the small mass differences between the Higgs
states in this regime and their relative large total decay widths, the
discrimination between the particles is challenging at the LHC and in some 
cases even impossible.  We propose to use the missing mass technique in the
Higgs--strahlung  process in $\ee$ collisions to distinguish between the two
CP--even Higgs eigenstates $h$ and $H$, relying on their $b\bar{b}$ decay in
the $b \bar b \ell^+ \ell^-$ event sample.  $Ah$ and $AH$ associated production
is then studied in the $4b$--jet event sample to probe the CP--odd $A$ boson.
At collider energies $\sqrt{s} \simeq$ 300 GeV and an integrated luminosity of
500 fb$^{-1}$, accuracies in the mass measurement of the CP--even Higgs bosons
are expected to range from 100 to 300 MeV, while for the CP--odd $A$ boson,
accuracies of less than 500 MeV can be obtained.  

\end{abstract}

\newpage

In the Minimal Supersymmetric Standard Model (MSSM), two Higgs doublets are
needed to break the electroweak symmetry and therefore, there are five physical
Higgs states: two CP--even Higgs particles $h$ and $H$, a CP--odd or
pseudoscalar Higgs boson $A$, and two charged Higgs particles $H^\pm$
\cite{HHG}. The intense--coupling regime \cite{intense,paper} is characterized
by a rather large value of the ratio of the vacuum expectation values of the
two doublet fields, $\tb =v_2/v_1 \gsim 10$, and a mass for the pseudoscalar
$A$ boson that is close to the maximal (minimal) value  of the CP--even $h$
($H$) boson mass.  In such a scenario, an almost mass degeneracy of the neutral
Higgs particles occurs, $M_h \sim M_A \sim M_H \sim 100$--140 GeV.  The
couplings of both the CP--even $h$ and $H$ particles to gauge bosons and
isospin up--type fermions are suppressed, and their couplings to down--type
fermions, in particular to $b$--quarks and $\tau$ leptons, are strongly
enhanced. The interactions of both Higgs particles therefore approach those of
the pseudoscalar Higgs boson which does not couple to massive gauge bosons as a
result of CP invariance, and for which the couplings to isospin $-\frac12 \,
(+\frac12)$ fermions are (inversely) proportional to $\tb$. Because of this
enhancement, the branching ratios of the $h$ and $H$ bosons into  $b\bar{b}$
and $\tau^+\tau^-$ final states are by far dominant, with values of $\sim 90$\%
and $\sim 10$\%,  respectively, similarly to the pseudoscalar Higgs case.  A
corollary of this feature is that the total decay widths of the three neutral
Higgs particles are rather large, being of the same order as the mass
differences.\s

As discussed in Ref.~\cite{paper}, this leads to a rather difficult situation
for the detection of these particles at the LHC. The branching ratios of the
interesting decays which allow the detection of the CP--even Higgs bosons,
namely $\gamma \gamma$, $WW^* \to \ell \ell \nu \nu$  and $ZZ^* \to  4\ell$,
are too small and prevent serious analyses. The $b\bar b$ decay mode has a too
large QCD background to be useful. For $\tau^+ \tau^-$ decays, the expected
experimental resolution on the invariant mass of the tau system is about 10--20
GeV and thus clearly too large for distinct Higgs particle observation; rather,
one would simply observe a relatively broad excess over the background,
corresponding to $A$ and $h$ and/or $H$ production. A way out, as suggested in
Ref.~\cite{paper}, is to rely on the decays into muon pairs with the Higgs
bosons produced in association with $b\bar b$ pairs, $gg/q\bar q \to b\bar b
+\Phi$ with $\Phi = h, H$ and $A$; see also Ref.~\cite{ggbbH}. Although the
decay is rare, BR($\Phi \to \mu^+\mu^-) \sim 3.3 \times 10^{-4}$,  the dimuon
mass resolution is expected to be as good as 1 GeV, i.e.  comparable to the
Higgs total widths for $M_\Phi \sim 130$ GeV\footnote{The Higgs--strahlung and
vector-boson fusion processes for the production of the $h$ and $H$ bosons, as
well as associated production of the three neutral Higgs particles with top
quarks, will have smaller cross sections than in the SM due to the suppressed
couplings of the particles involved.  The production of the three Higgs
particles in the gluon--gluon process, $gg \to \Phi \to \mu^+ \mu^-$, although
bearing large rates will suffer from the huge Drell--Yan $pp \to \gamma^*, Z^*
\to \mu^+\mu^-$ background process \cite{paper}.}. However, even in this case,
it is possible to resolve only two Higgs peaks in favorable situations. In
general, the detection of the three individual Higgs bosons is very
challenging, and in some cases even impossible at the LHC\footnote{An
alternative possibility at the LHC is diffractive Higgs production
\cite{khoze} where, based on the recoil mass technique, very good proton beam
energy resolution and precise luminosity measurements are crucial to resolve
the Higgs signals and perform accurate mass determinations.}.\s


In $\ee$ collisions \cite{ee-xs}, the CP--even Higgs bosons can be produced in
the Higgs--strahlung, $\ee \to Z+h/H$, and in the vector-boson fusion,
$\ee \to \nu \bar{\nu}+h/H$, processes. The CP--odd particle cannot be probed
in these channels due to its zero-couplings to gauge bosons at tree level,
but it can be produced in association with the $h$ or $H$ bosons in the
reactions $\ee \ra A + h/H$. Earlier studies \cite{TESLA} indicated that
the vector boson fusion processes are difficult to use in this context,
as the full final state cannot be reconstructed. In turn, the Higgs--strahlung
and the Higgs pair production processes, as will be demonstrated in this
note, have a great potential to explore the individual $h,H$ and $A$
states in the intense--coupling regime and to allow the measurement of 
their masses.\s 

The cross sections for the Higgs--strahlung and pair production processes are 
mutually complementary coming either with a coefficient $\sin^2(\beta- \alpha)$
or $\cos^2(\beta -\alpha)$, with $\alpha$ being the mixing angle in the 
CP--even Higgs sector: 
\beq
\sigma(\ee \to Z+h/H) &=& \sin^2/\cos^2 (\beta-\alpha) \sigma_{\rm SM} 
\non \\ 
\sigma(\ee \to A+h/H) &=& \cos^2/\sin^2 (\beta-\alpha) \bar{\lambda} 
\sigma_{\rm SM} \non 
\eeq
where $\sigma_{\rm SM}$ is the SM Higgs cross section in the strahlung process
and $\bar{\lambda}$ $\sim$1 for $\sqrt{s} \gg M_A$ accounts for P--wave
suppression near the kinematical threshold for the production of two spin--zero
particles.  Since $\sigma_{\rm SM}$ is rather large, being of the order of
50--100 fb for a Higgs boson with a mass $\sim130$ GeV at a c.m. energy
$\sqrt{s} \sim 300$--500 GeV, the production and the detection of the three
neutral Higgs bosons should be straightforward for an integrated luminosity of
$\int \! {\cal L} \sim 0.5$--1 ab$^{-1}$, as expected at future linear $\ee$
colliders such as TESLA \cite{TESLA}.\s

In Fig.~1, the production cross sections for the Higgs--strahlung and Higgs
pair production of the neutral Higgs particles are shown as a function of the
c.m. energy. We have chosen the same three representative scenarios P1, P2 and
P3 discussed in Ref.~\cite{paper}: $\tb=30$ and $M_A=125, 130$ and $135$ GeV. 
The maximal mixing scenario where the trilinear Higgs--stop coupling is given
by $A_t \simeq \sqrt{6} M_S$ with the common stop masses fixed to $M_S=1$ TeV
has been adopted; the other SUSY parameter will play only a minor role and have
been set to 1 TeV, while the top quark mass is fixed\footnote{We have preferred
to use this value instead of the recent Tevatron central value of $m_t=178$ GeV
to allow for a comparison with the analysis performed for the LHC in
Ref.~\cite{paper}.} to $m_t=175$ GeV. The resulting Higgs masses, couplings and
branching ratios shown in Table 1 have been obtained using the program {\tt
HDECAY} \cite{hdecay} in which the routine {\tt FeynHiggsFast} \cite{feynhiggs}
is used for the implementation of the radiative corrections. As apparent from
Fig.~1, values of $\sqrt{s}$ not too far above the kinematical thresholds of
these reactions are favored within our scenarios with $M_\Phi \sim 130$ GeV,
since the cross sections scale like $1/s$ as the processes are mediated by
$s$--channel gauge boson exchange. We will thus choose to operate the $\ee$
collider at $\sqrt{s}=300$ GeV in the present analysis, as the production cross
sections are large enough for all cases considered. \s

\begin{table}[h!]
\renewcommand{\arraystretch}{1.1}
\begin{center}
\vspace*{-2mm}
\begin{tabular}{|c|c||c|c||c|c|} \hline
Point & $\ \ \Phi \ \ $ & $\ \ M_\Phi \ \ $  & $\ \ \Gamma_\Phi \ \ $ &
\ \ BR$(b \bar b)$ \ \ & BR($\tau^+ \tau^-$) \\ \hline
   &$h$ & 123.3 & 2.14 & $ 0.905$ & $0.093$ \\
P1&$A$ & 125.0 & 2.51 & $ 0.905$ & $0.093$ \\
   &$H$ & 134.3 & 0.36 & $0.900$ & $0.094$ \\
\hline
   &$h$ & 127.2 & 1.73 & $0.904$ & $0.093$ \\
P2&$A$ & 130.0 & 2.59 & $0.904$ & $0.094$ \\
   &$H$ & 135.5 & 0.85 & $0.900$ & $0.094$ \\
\hline
   &$h$ & 129.8 & 0.97 & $0.903$ & $0.094$ \\
P3&$A$ & 135.0 & 2.67 & $0.904$ & $0.094$ \\
   &$H$ & 137.9 & 1.69 & $0.902$ & $0.095$ \\ \hline
\end{tabular}
\end{center}
\vspace*{-3mm}
\caption[]{\it Masses, total decay widths (in GeV) and some decay branching 
ratios of the MSSM neutral Higgs bosons for the points P1, P2 and P3 with
$\tb=30$}
\vspace*{-5mm}
\end{table}

\begin{figure}[h!]
\begin{center}
\centerline{\epsfig{file=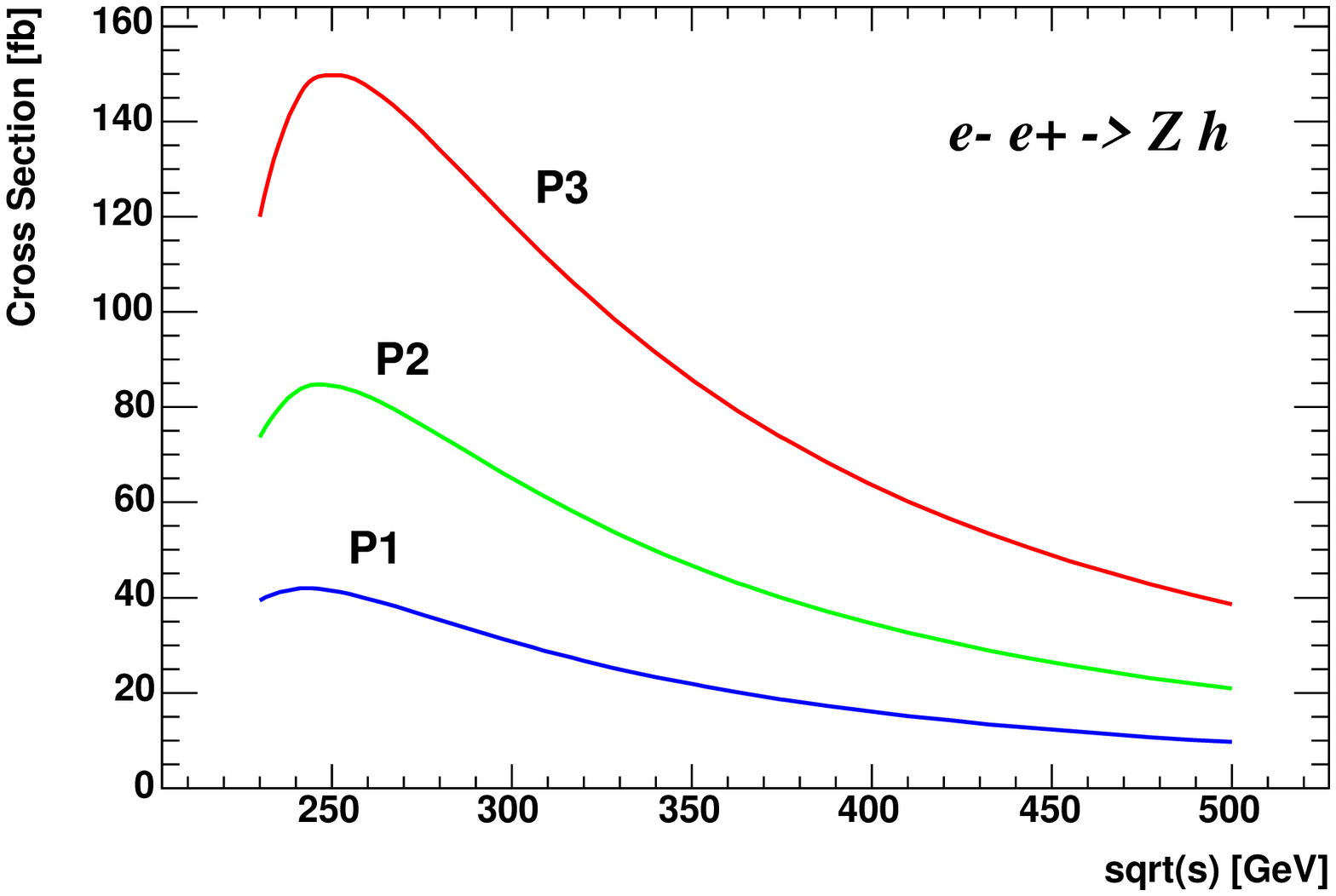,width=8.3cm,height=6.5cm}
\epsfig{file=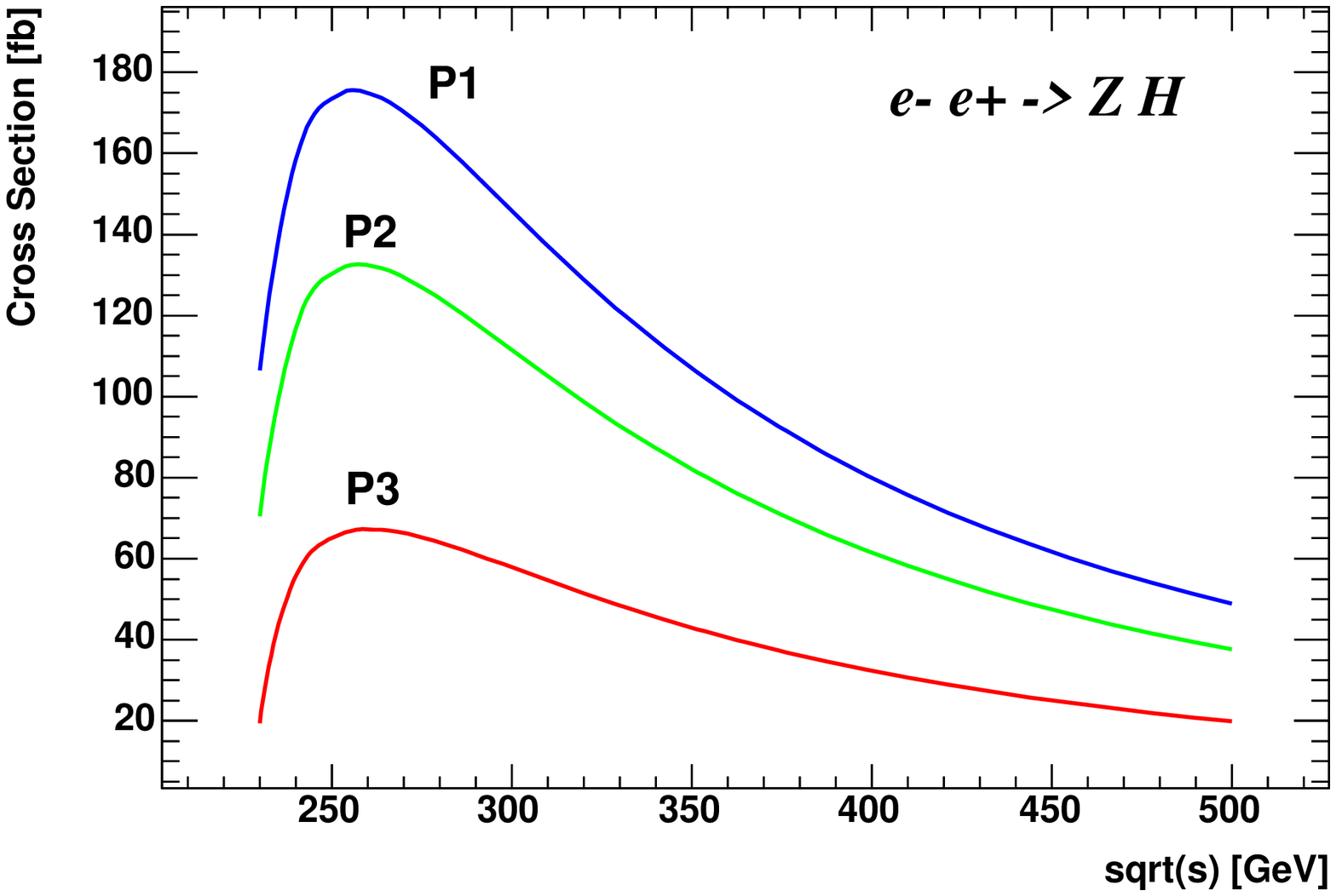,width=8.3cm,,height=6.5cm}}
\centerline{\epsfig{file=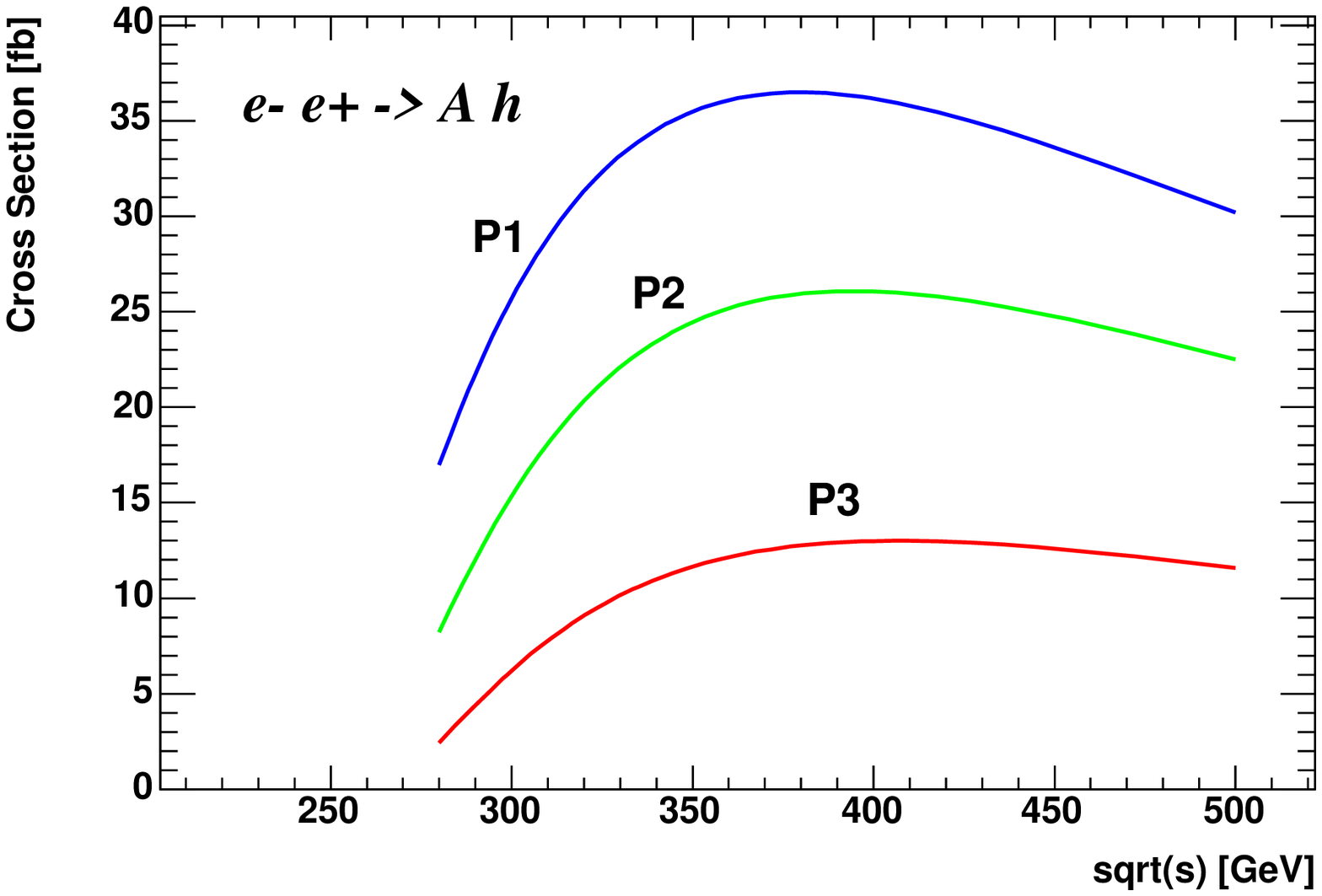,width=8.3cm,height=6.5cm}
\epsfig{file=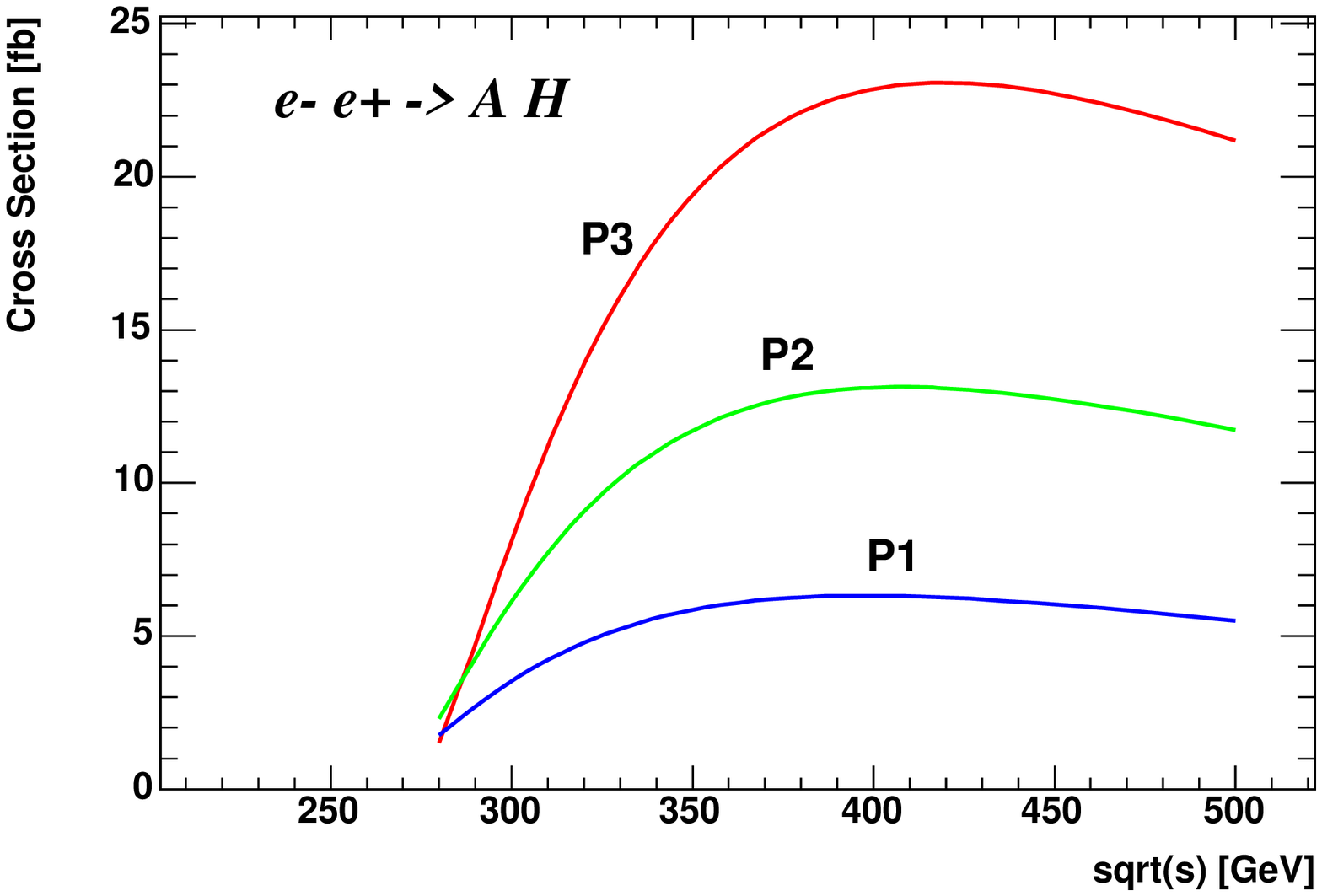,width=8.3cm,height=6.5cm}}
\vspace*{-.3cm}
\caption{\it The production cross sections for the Higgs--strahlung (upper 
plots) and Higgs pair processes (lower plots) for the MSSM parameter points P1,
P2, and P3 with $M_A = 125, 130, 135$ GeV and $\tb =30$ in the maximal 
mixing scenario.}
\end{center}
\vspace*{-1.2cm}
\end{figure}

The Higgs--strahlung processes offers the most promising  way to discriminate
between the two CP--even Higgs particles, since the pseudoscalar boson $A$ is
not  involved. For the SM Higgs boson, as was widely demonstrated, the
recoil mass technique in both leptonic and hadronic $Z$ decays allow very
precise determination of its mass; for instance an accuracy of $\sim 40$ MeV
for a mass of $\sim 120$ GeV can be achieved \cite{TESLA}.  In the
intense--coupling scenario, where the two scalar $h$ and $H$ bosons are close
in mass and are often produced with different rates, some of them being small,
the impact of initial state radiation (ISR) and beamstrahlung is important and
should be carefully taken into account. We have performed a detailed
simulation, including the signal and all the main background reactions using
the program package {\tt CompHEP} \cite{comphep} interfaced \cite{interface}
with {\tt PYTHIA} \cite{PYTHIA}, as well as a simulation of the detector
response with the code {\tt SIMDET} \cite{SIMDET}. The analysis reveals that
the most promising way for measuring the $h$ and $H$ boson masses is to select
first the $\ell^+ \ell^- b \bar{b}$ event sample ($\ell=e/\mu$), followed by
the recoil $Z$ mass technique. However, without cuts and $b$--quark tagging,
the signals from the $h$ and $H$ bosons cannot be resolved, as illustrated in
Fig.~2 in the case of the parameter point P1.\s

\begin{figure}[!ht]
\begin{center}
\vspace*{-.9cm}
\centerline{\epsfig{file=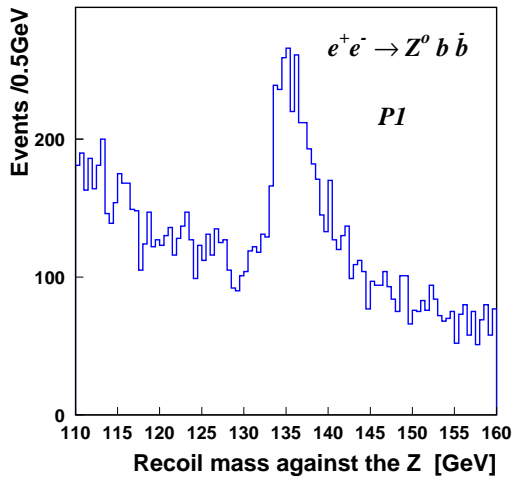,width=7cm}}
\end{center}
\vspace*{-.9cm}
\caption{\it The recoil mass distribution for signal and background 
including ISR, beamstrahlung and detector smearing for the parameter point P1
before cuts and $b$--tagging.}
\vspace*{-.3cm}
\end{figure}

If some realistic $b$--tagging is applied and the surviving $b \bar{b} \ell^+
\ell^-$ events are required to pass the following cuts: $i)$ the dilepton
invariant mass is within $M_{\ell^+ \ell^-} = 90 \pm 6$ GeV, i.e. compatible
with the $Z$ boson, $ii)$ each jet
energy has $E_{j} \geq 12$ GeV, $iii)$ the angle between two jets is
$\angle(j_1,j_2) \geq 95$ degrees, the separation of the two Higgs signal peaks
is possible and the masses are accessible. Simulation results for the case of
TESLA, as an example, and for the MSSM parameter points P1, P2 and
P3 are shown in Fig.~3.  The selection efficiencies are found to be 68\% for
the signal reaction, while they are at the level of 22\% for the $\ell^+
\ell^-b \bar b$, 6.4\% for the $\ell^+ \ell^-c\bar c$ and 0.1\% for the
$\ell^+\ell^-q \bar q$ $(q=u,d,s)$ background processes.\s
                                                                                 
\begin{figure}[h!]
\begin{center}
\vspace*{-0.9cm}
\centerline{\psfig{file=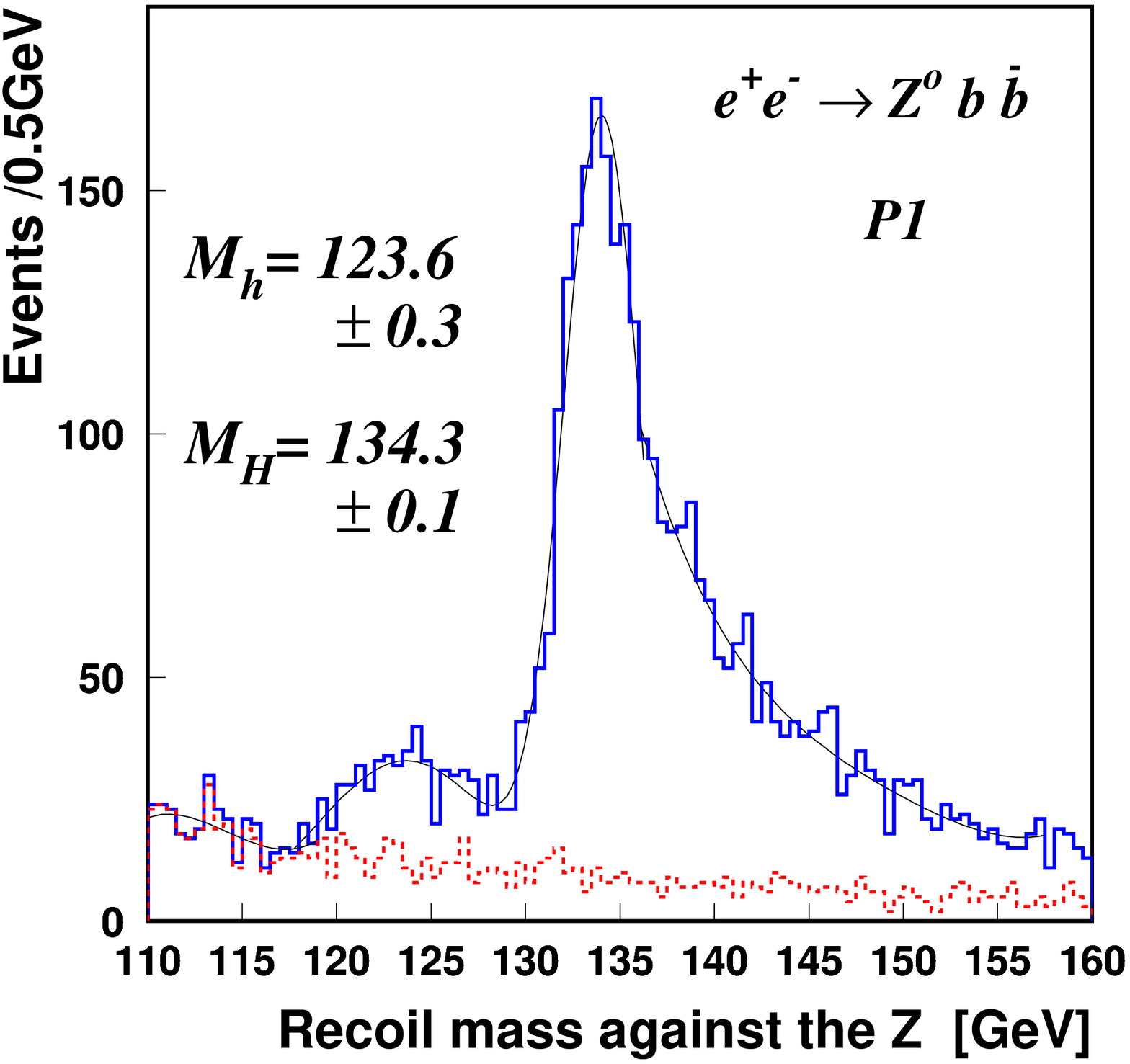,width=7.3cm}}
\centerline{
\psfig{file=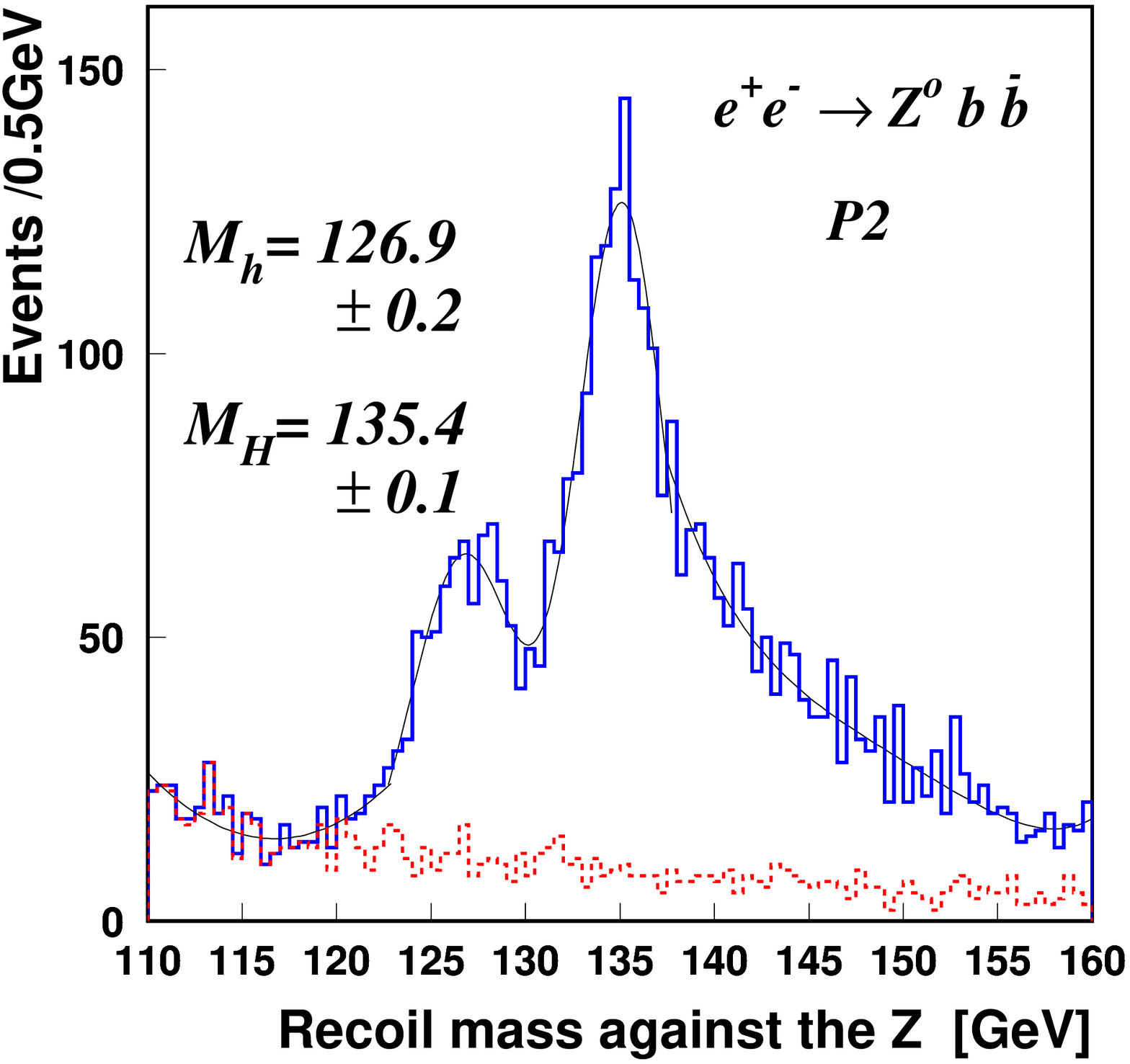,width=7.3cm}
\psfig{file=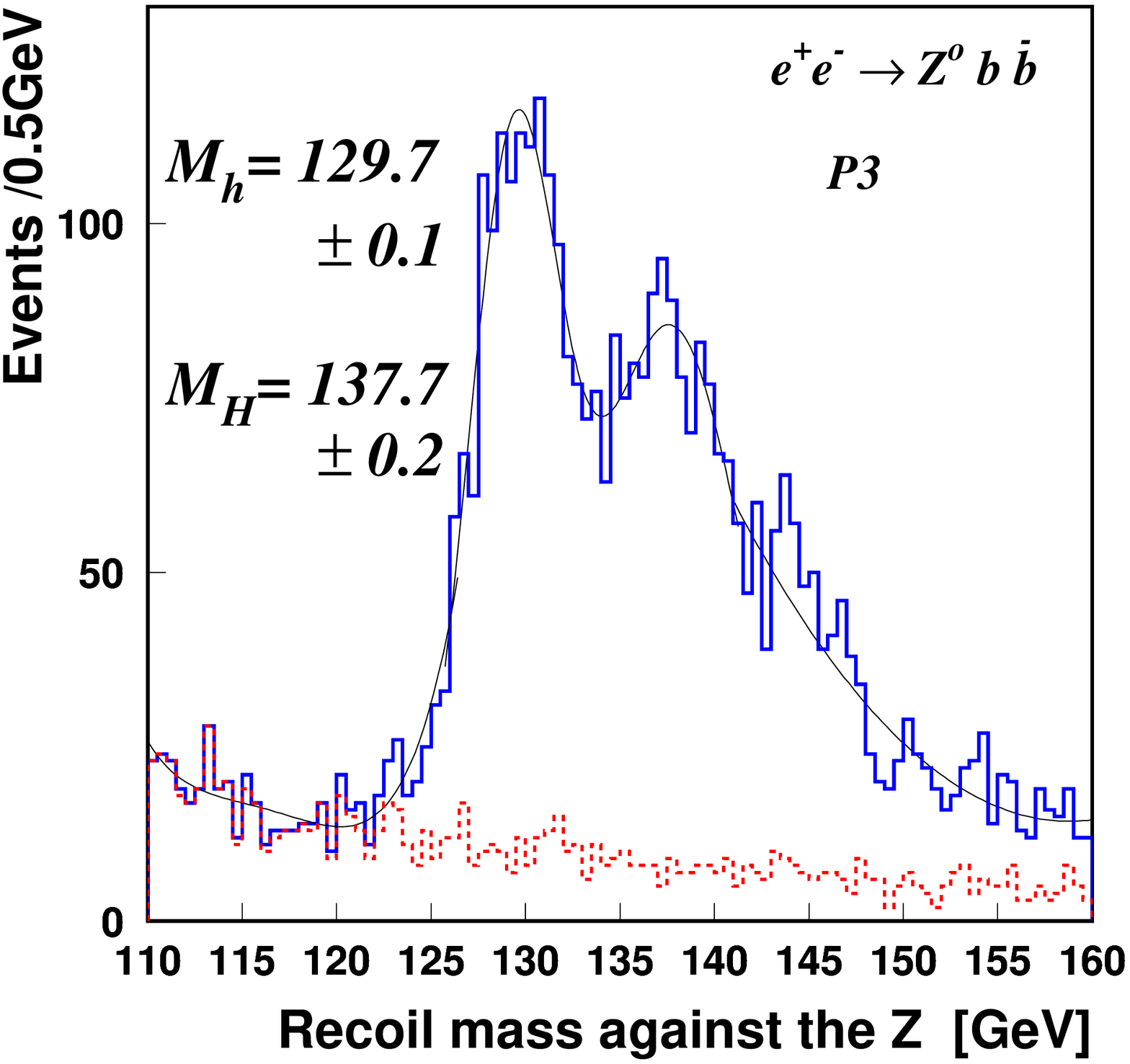,width=7.3cm}}
\end{center}
\vspace*{-.99cm}
\caption{\it The recoil mass distributions for the sum of signal and
background including ISR, beamstrahlung and detector smearing for the 
parameter points P1, P2, P3 after cuts and $b$--tagging. The background is
separately shown as dashed histogram. 
The solid line is the result of a fit, with values for $M_h$ and $M_H$ as indicated.}
\vspace*{-.4cm}
\end{figure}

As evident from Fig.~3,  the masses of the $h$ and $H$ particles can be
determined with accuracies of the order of 100--300 MeV at a 300 GeV collider
energy and  with 500 fb$^{-1}$ accumulated luminosity.  Such uncertainties in
the mass measurements are significantly smaller than the typical mass
differences between the two Higgs states. They are however larger
than the corresponding accuracy for the SM Higgs boson. At higher c.m. 
energies, the mass determination will be significantly worse as a
consequence of the smaller production cross sections, degraded energy
resolution of the more energetic leptons and the stronger impact of ISR and
beamstrahlung. It would become very difficult to resolve the $h$ and $H$
signals as is demonstrated in Fig.~4 for $\sqrt{s}$ = 500 GeV and two
times larger integrated luminosity of $\int {\cal L}=1$ ab$^{-1}$.
Here, only the Higgs signal events are shown.\s

\begin{figure}[htbp]
\begin{center}
\vspace*{-0.5cm}
\centerline{\psfig{file=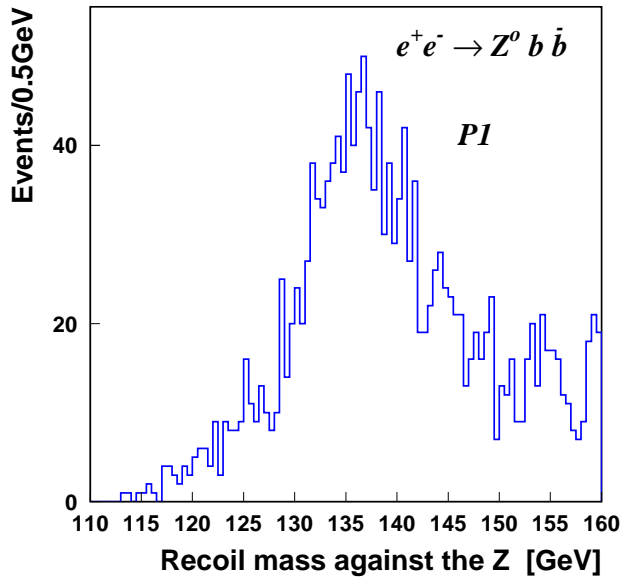,width=8.5cm}}
\end{center}
\vspace*{-1.cm}
\caption{\it  The recoil mass distribution for signal events 
in $\ee \to Zb\bar b$ at $\sqrt s = 500$ GeV and $\int {\cal L}=1$ ab$^{-1}$
for the parameter point P1.}
\vspace*{-.15cm}
\end{figure}

Once the $h$ and $H$ boson masses are known from the recoil mass technique,
attention should be directed to the mass determination of the $A$ particle
which can be probed in the complementary pair production channels $\ee \to
A+h/H$. This can be achieved either via the reconstruction of the $b\bar{b}$
and/or $\tau^+ \tau^-$ invariant masses or through a threshold scan. The first
method has been  discussed in Ref.~\cite{HA} for the production of heavier
Higgs bosons in the decoupling limit $M_A \sim M_H \gg M_Z$, in the reaction
$\ee \to HA \to 4b$ at $\sqrt{s}=800$ GeV. Accuracies of about 100 MeV for the
$H/A$ masses were obtained sufficiently above the reaction thresholds using the
dominant $b\bar bb\bar b$ and $b\bar{b} \tau^+ \tau^-$ final states \cite{desch}.\s  

In the intense--coupling regime, the three neutral Higgs bosons contribute to
the $b\bar{b}b\bar{b}$ and $b\bar{b}\tau^+ \tau^-$ final state signals. Since
typical $b$--jet energy resolutions are close to or somewhat larger than the
Higgs mass differences, it is challenging to discriminate between the $A
\rightarrow b\bar{b}$ and the $h/H \rightarrow b\bar{b}$ decays, as illustrated
in Fig.~5 where all possible $b\bar b$ mass combinations for the signal and the
sum of signal and background events are shown for the parameter point P1
as an illustration.\s

\begin{figure}[htbp]
\begin{center}
\vspace*{-0.8cm}
\centerline{\psfig{file=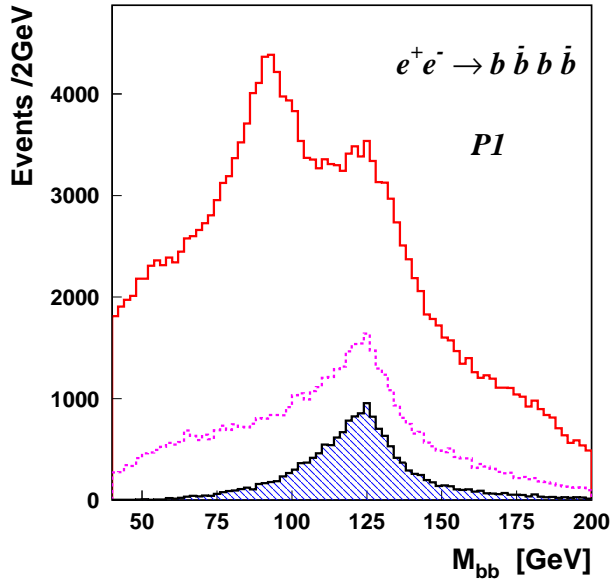,width=8.5cm}}
\end{center}
\vspace*{-1.15cm}
\caption{\it  $b\bar{b}$ invariant mass distributions for the combined signal 
and total background (top), for the sum of signal and combinatorial
background (dashed) and only the signal (shaded histogram)
 for the parameter point P1.}
\vspace*{-0.35cm}
\end{figure}

To associate the correct $b\bar{b}$ mass combination to the $A$ and $h/H$ boson
decay into $4b$ final state events, we use the ``combinatorial mass 
difference" method. After selecting $4b$--jet events by means of $b$--tagging, 
we consider all three possible combinations of 2 $b$--jet pairs. Only one of 
them is the ``physical" combination where both $b$-quarks in each pair 
correspond to one of the decaying Higgs particles, while the other two 
combinations are combinatorial background. \s

Because of the well defined kinematics in the process $e^+e^- \rightarrow A+h/H 
\rightarrow b\bar b \, b\bar b$, the angle between two $b$--jets in the Higgs
decay
\beq
\angle (b\bar b)\simeq 2 \times{\rm arctan}\left(2 \times \sqrt{\frac{M_{\Phi}^2-
4m_{b}^2} {s-4M_{\Phi}^2}} \right)\ \hspace{15mm} (\Phi = A, h, H)
\eeq
is about $115^\circ$ for our parameter set and independent of the Higgs 
particles since their masses are almost degenerate. The influence of ISR and 
beamstrahlung leads to some smearing of the corresponding angular distribution
as shown in the left--hand side of Fig.~6. A sharp distribution is evident for 
the ``correct" or signal $b$--jet pair, while the combinatorial $b$--jet 
background pairing leads to a flat distribution.\s  

In addition, $b$--jet pairs originating from Higgs decays are more centrally
produced than the combinatorial background, as evident from the right--hand side
of Fig.~6. Thus, the separation of the ``physical" combinations from the
combinatorial background might be achieved by means of the following cuts: $i)$
$-0.95<\cos\theta_{ b_1 b_2}<-0.3$ and $ii)$ $|\cos\theta_{bb-\rm pair} |
<0.7$, where $\theta_{b_1 b_2}$ is the angle between two $b$--jets and
$\theta_{bb-\rm pair}$ the polar angle of the $b \bar b$ system.  The
``physical pairs" are selected with an efficiency of about 85\%, whereas the
background combinations are selected with an efficiency of about 20\%.\s

\begin{figure}[htbp]
\begin{center}
\vspace*{-0.2cm}
\centerline{\psfig{file=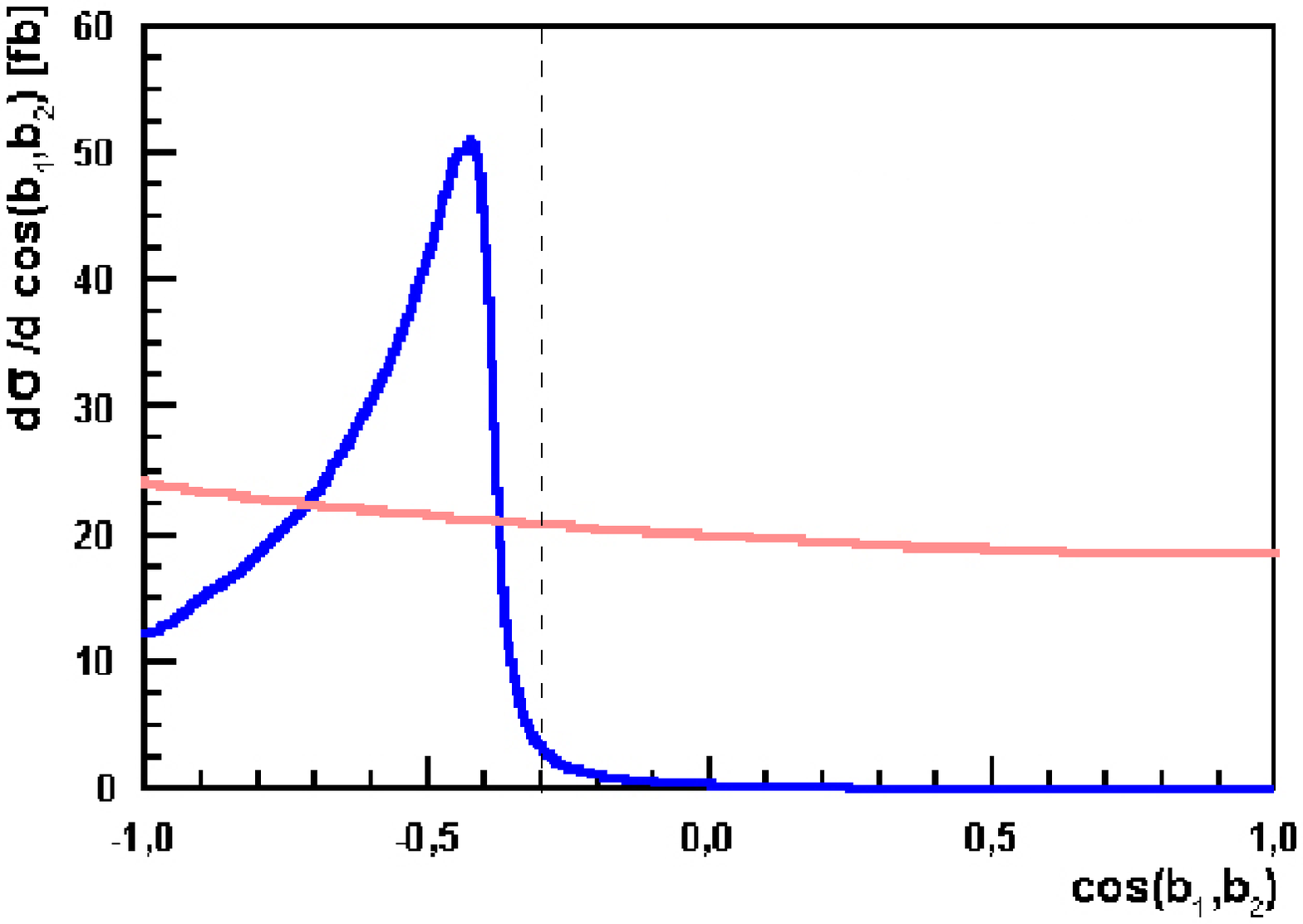,width=8.5cm,height=7.99cm}\hspace*{-3mm}
\psfig{file=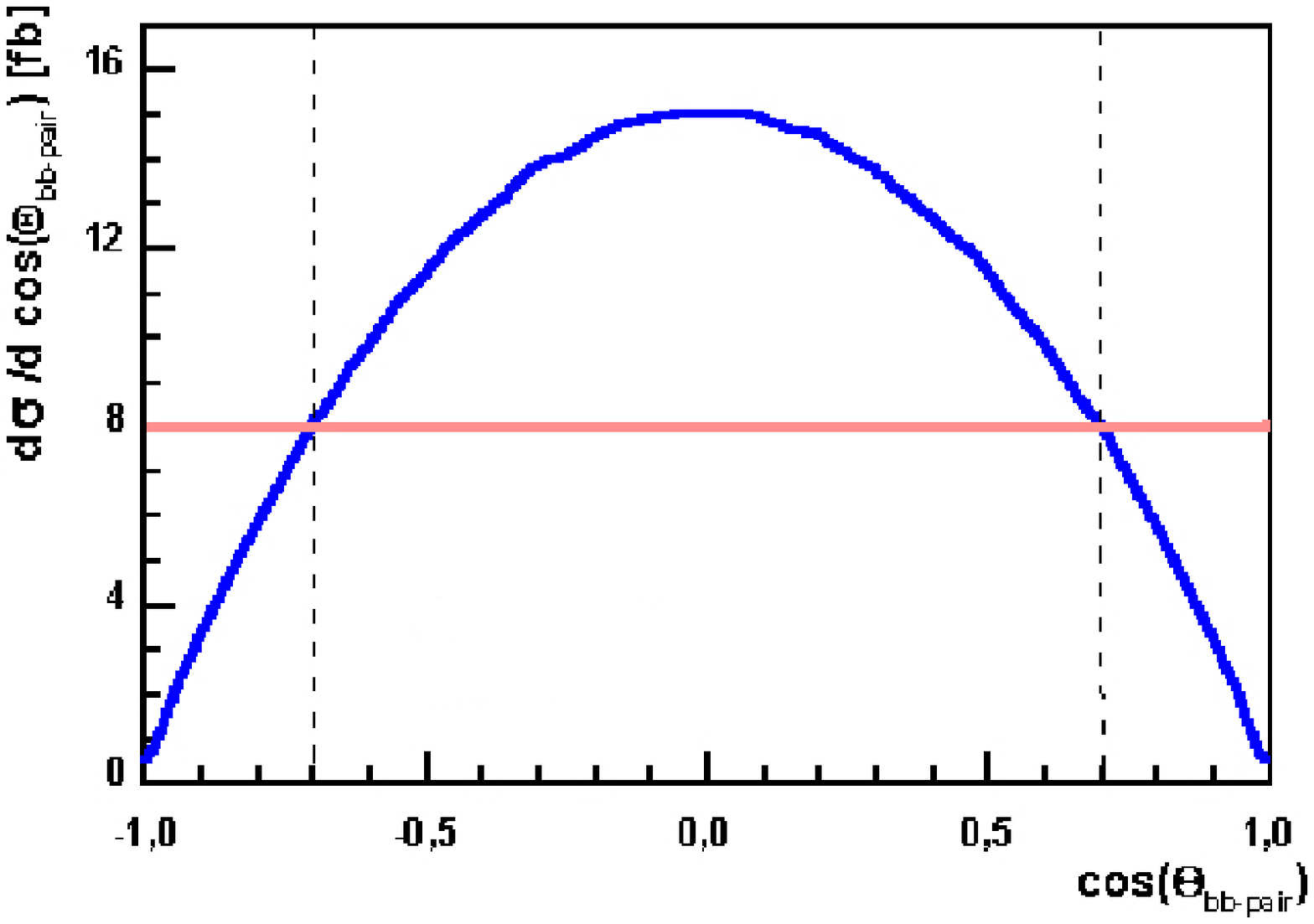,width=8.5cm,height=7.99cm}}
\end{center}
\vspace*{-1.5cm}
\caption{\it  The differential distributions of the angle between two $b$--jets
for the ``physical" and combinatorial background combinations (left) and 
the polar angular distribution for $b$--jet pairs (right). The background 
concerns the flat distributions in either case.}
\vspace*{-0.15cm}
\end{figure}

In the discrimination between $A h$ and $A H$ production, we use the average
mass, $\overline{M}=\frac{1}{2}(M_{1}+M_{2})$, where $M_{1}$ and $M_{2}$
are randomly chosen among two invariant masses of 2 $b$--jets in the 
event sample after cuts. If $\overline{M}$ is closer to the known mass of the 
$h$ boson measured by the recoil mass technique, we associate the $4b$--jet 
final state to $\ee \to A h$ production, otherwise it is associated to $\ee \to
A H$.\s

Finally, the selection of the pseudoscalar boson $A$ from the $A h$ and $A H$
pairing relies on some probability estimation based on the function 
\beq 
{\cal P}=\frac12+\frac12 \times \frac{M_{2}-M_{1}}{M_{2}+M_{1}} 
\eeq 
If, for instance, the former step favors $Ah$ production for a particular 
$4b$--jet event, the function ${\cal P}$ gives the probability that the first 
chosen invariant mass $M_{1}$ is the mass of the $h$ boson.  This probability 
value is compared with a uniformly distributed random number $r$ in the range 
$[0,1]$. If ${\cal P}>r$, the association $M_{h}=M_{1}$ and $M_{A}=M_{2}$ is 
performed, whereas for the opposite case ${\cal P} <r$, we assign $M_{h}=M_{2}$
and $M_{A}=M_{1}$.  The same procedure is applied if $A H$ pair production has
been favored in the first step. Resulting $b \bar b$ mass spectra for the MSSM
parameter points P1, P2 and P3 are shown in Fig.~7. Only the $2b$--jet masses
which have been assigned to the pseudoscalar $A$ boson are displayed, and all 
$4b$--jet background sources have been taken into account. \s

\begin{figure}[h!]
\begin{center} 
\vspace*{-0.5cm}
\centerline{\psfig{file=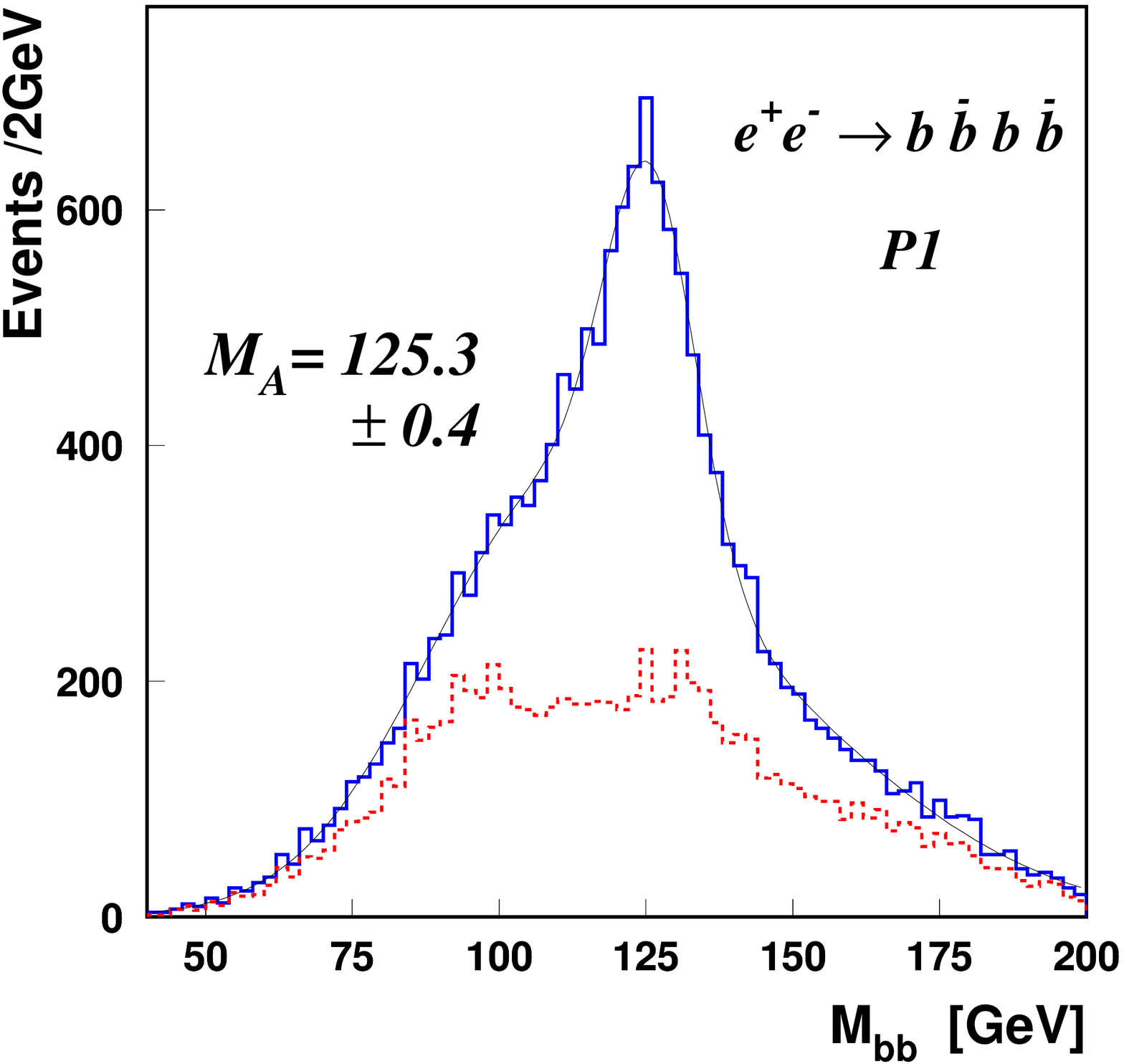,width=7.7cm}}
\centerline{
\psfig{file=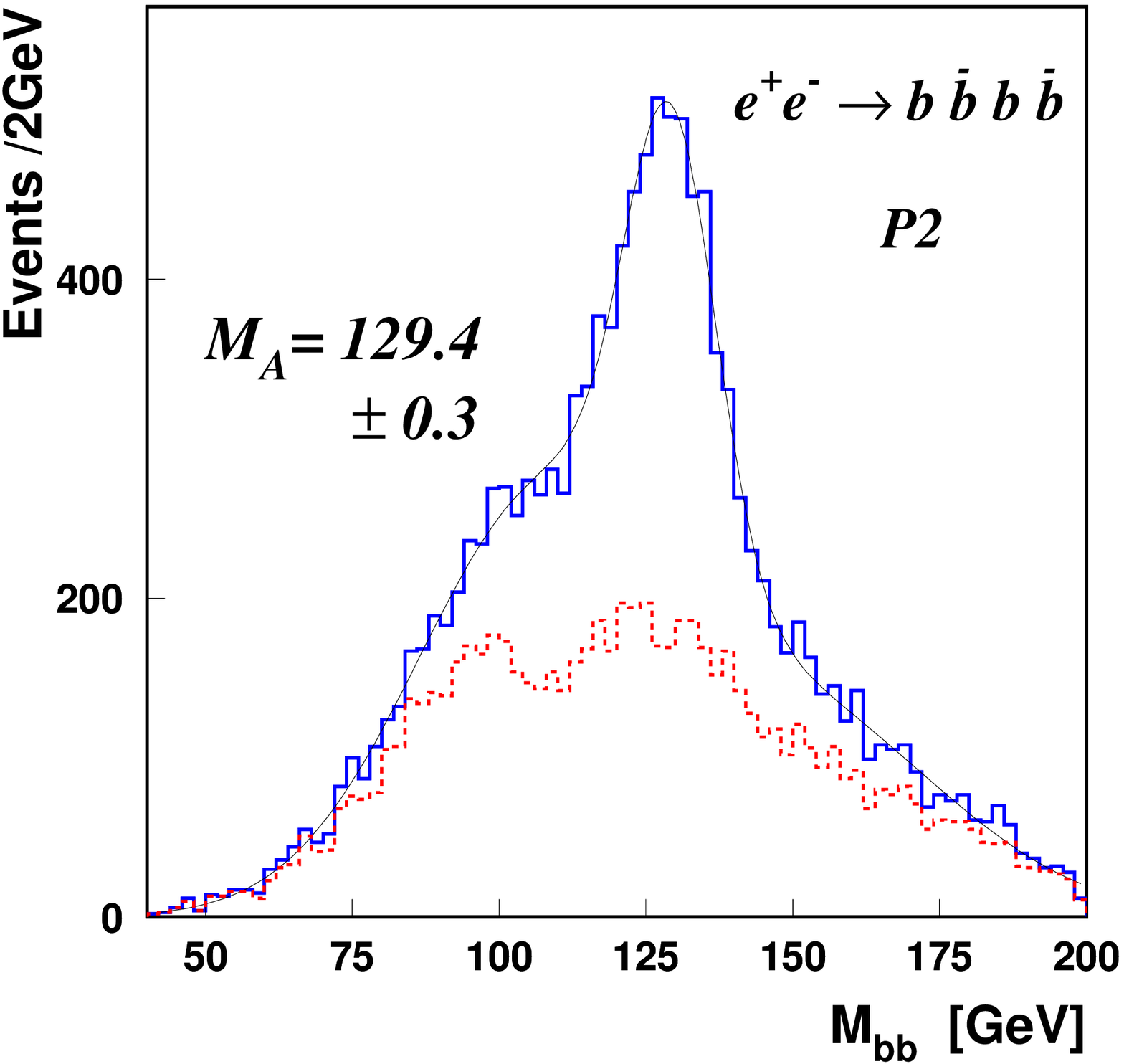,width=7.7cm}
\psfig{file=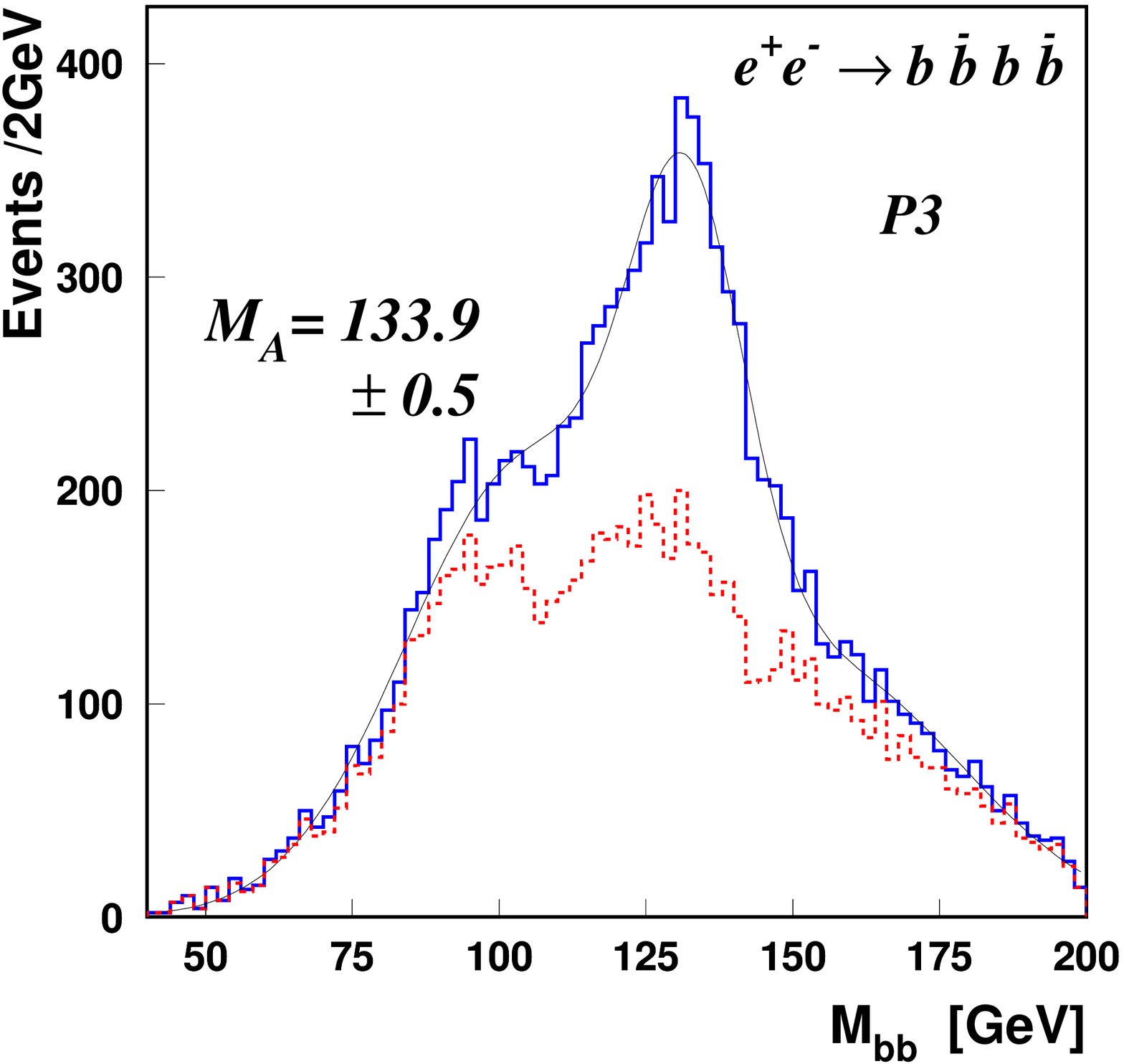,width=7.7cm}}
\vspace*{-.5cm}
\end{center}
\caption{\it The two $b$--jet invariant mass associated to the pseudoscalar 
$A$ boson after cuts and the selection procedure for the parameter points P1, 
P2 and P3. The top histogram represents signal plus background,
while the dashed histogram the background only. 
The solid line is the result of a fit, with values for $M_A$ as indicated.}
\vspace*{-.3cm}
\end{figure}

As fits to these histograms revealed, the mass of the pseudoscalar Higgs boson 
$A$ can be measured with an accuracy of 300 to 500 MeV, once the measured 
masses of the $h$ and $H$ particles with their corresponding errors have been
taken into account. Such experimental accuracies, although larger than those 
for the SM Higgs boson, are smaller than the typical mass differences $M_A-M_h$
or $M_H- M_A$ in the chosen scenarios.\s  
       
One may expect a more precise determination of the mass values for the $A$
boson by measuring the $\ee \to A+h/H$ production cross sections near the
respective kinematical thresholds, where the cross sections rise as $\sigma
\sim \beta^3$ with $\beta \sim \sqrt{1- 4 M_A^2/s}$. This is very similar to
scalar lepton pair production in $\ee$ collisions, $\ee \to \tilde \ell \tilde
\ell$ in supersymmetric models,  which has many common characteristics with the
process discussed here. Indeed, it has been shown \cite{slep} that slepton
masses of the order of 100 GeV can be measured with an accuracy of less than
0.1\% in a threshold scan.  Whether this holds also true for $\ee \to A+h/H$
production in the intense--coupling regime [the production cross sections are
smaller but the final states are cleaner] has to be studied in detail,
including ISR and beamstrahlung. This study is however beyond the scope of
this note. \s

In conclusion, the intense--coupling regime in the MSSM Higgs sector, in which
$\tb$ is rather large and the three neutral $h$, $H$ and $A$ Higgs particles
have comparable masses, is a difficult scenario to be resolved completely at
the LHC. In $\ee$ collisions, thanks to the clean environment and to the
complementarity of the available production channels, the separation of the
three states is possible. The Higgs--strahlung processes allows first to probe
the $h$ and $H$ bosons and to measure their masses from the recoiling mass
spectrum against the $Z$ boson; the best results are obtained by selecting the 
$b\bar b + \ell^+ \ell^-$ event sample and imposing $b$--jet tagging.  Then,
associated CP--even and CP--odd Higgs production would allow to probe the
pseudoscalar $A$ boson by direct reconstruction of its decay products. At
collider energies $\sqrt{s} \simeq$ 300 GeV and with integrated luminosities of
500 fb$^{-1}$, accuracies for the measurement  of the masses of the three
neutral Higgs particles are expected to range from 100 to 500 MeV, which is
smaller than the typical Higgs mass differences in this scheme. \s

In the study of the intense--coupling regime, the interplay between the LHC and
a future linear collider might be very important: on the one hand, any broad
peak information observed at the LHC will assist the choice of the appropriate
energy at the LC and on the other hand, characteristics of the Higgs states as
measured at the linear collider could constrain techniques to access further
observables at the LHC such as the gluon-gluon-Higgs couplings and the $\Phi
\to \mu^+\mu^-$ branching ratios, as the processes $gg \to \Phi \to \mu^+
\mu^-$ might be then possible to detect, a posteriori.

\subsubsection*{Acknowledgments:}

The work of A.D. is supported by the Euro--GDR Supersym\'etrie and by European
Union under contract HPRN-CT-200-00149. The work of E.B. and V.B.  is partly
supported by RFBR~04-02-16476, RFBR~04-02-17448, University of Russia 
UR.02.03.028, and Russian Ministry of Education and Science NS.1685.2003.2
grants. E.B. thanks the Fermilab 
Theoretical  Physics Department for the kind hospitality. V.B. 
acknowledges the warm hospitality of DESY Zeuthen.

\end{document}